\documentclass[runningheads]{llncs}
\usepackage{graphicx}

\usepackage[utf8]{inputenc}
\usepackage[hidelinks]{hyperref}

\usepackage{booktabs}
\usepackage{makecell} 

\begin{document}


\title{A CNN toolbox for skin cancer classification}
\author{Fabrizio Nunnari \and Daniel Sonntag }

\authorrunning{F. Nunnari, D. Sonntag}

\institute{Deutsches Forschungszentrum für Künstliche Intelligenz (DFKI)\\ Technical Report \\
\email{\{fabrizio.nunnari,daniel.sonntag\}@dfki.de}\\
\url{http://iml.dfki.de/}}

\maketitle

\begin{abstract}
We describe a software toolbox for the configuration of deep neural networks in the domain of skin cancer classification.
The implemented software architecture allows developers to quickly set up new convolutional neural network (CNN) architectures and hyper-parameter configurations. At the same time, the user interface, manageable as a simple spreadsheet, allows non-technical users to explore different configuration settings that need to be explored when switching to different data sets. In future versions, meta leaning frameworks can be added, or AutoML systems that continuously improve over time. Preliminary results, conducted with two CNNs in the context melanoma detection on dermoscopic images, quantify the impact of image augmentation, image resolution, and rescaling filter on the overall detection performance and training time.

\end{abstract}

\section{Introduction}

The skin cancer death rate has escalated sharply in the USA, Europe and Australia \cite{celebi_dermoscopy_2019}. However, with proper early detection, the survival rate after surgery (wide excision) reaches 98\%.
For this reason, the research community has put a significant effort in the early detection of skin cancer through the inspection of images \cite{masood_computer_2013}. Recently, the best results has been achieved using transfer learning on Convolutional Neural Networks (e.g., \cite{esteva_dermatologist-level_2017,haenssle_man_2018,fujisawa_deep-learning-based_2018}).

As reported by Brinker et al. \cite{brinker_skin_2018}, regardless of the similarities in terms of sensitivity, specificity, and ROC AUC, those works are hardly comparable with each other because they are all based on different datasets (often proprietary), and use different CNN architectures and hyper-parameters.

Hence, given a new dataset, characterised for example by its own resolution, settings (lenses and light conditions), type (dermoscopic or clinical), and ethnicity (caucasian, asiatic, worldwide), choosing for the best CNN architecture and hyper-parameters is not straightforward. For example, one of the mostly recent influencing works (Esteva et. al \cite{esteva_dermatologist-level_2017}) showed a CNN that matches the accuracy of expert dermatologists when trained on more than 126k images. However, Fujisawa et al. \cite{fujisawa_deep-learning-based_2018} showed that, with a higher image augmentation (24x) and image resolution (1k), the same performances can be achieved using less than 5000 images. This is very important to the general area under study with less data material. 

Also, we have reports from the pre-CNN era, when features were extracted manually and image pre-processing was required, about the importance of extended segmentation \cite{burdick_rethinking_2018} and color filtering \cite{barata_two_2014} to improve the performance of classification. However, such techniques have not been applied in conjunction with deep learning approaches, where performance gains are mostly pursued by increasing the size of the training sets.

We try to enhance the infrastructure for research based on previous implementations \cite{DBLP:journals/corr/abs-1803-04818,DBLP:journals/corr/abs-1709-01476} to explore a number of options that require a considerable amount of software development. To address the two above-mentioned issues (lack of cross-CNN comparison and lack of integration of pre-processing techniques) we develop a software platform for the easy and systematic exploration of (hyper-)parameters governing the performances of image augmentation and CNNs.

Our platform (which will be released as open-source software once out of beta stage) has two target user groups:
\emph{Developers}, who need a structured and extensible software architecture to experiment with new image processing techniques and CNN architectures, and
\emph{Practitioners} in the field of dermatology, who do not necessarily have the competencies to script new software, but do need to explore the performances of existing techniques when new datasets become available.


%
%
%
\section{Training and Testing Pipeline}

\begin{figure}[t]
    \centering
    \includegraphics[width=\textwidth]{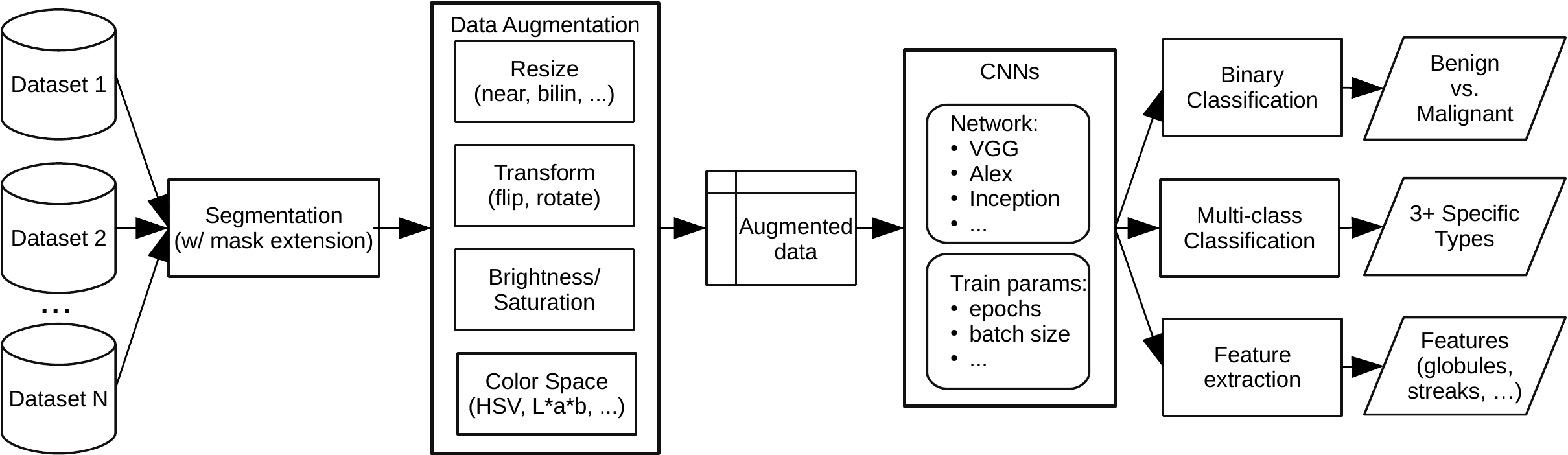}
    \caption{Training/Testing pipeline of the proposed architecture}
    \label{fig:pipeline}
\end{figure}

Figure \ref{fig:pipeline} shows the pipeline for the training and testing procedure of the proposed architecture. Starting from left,  a dataset is chosen, and all images are (optionally) segmented and the mask is extended.
The segmented images then go through an augmentation procedure, which includes the possibility to resize (using different filters), transform (flip/rotate), modulate brightness and saturation, and change the color space. Images are augmented on-the-fly.
The augmented images are then sent to a CNN which can output a binary classification (e.g., malignant vs. benign lesion), a multi-class probability distribution, or a pixel-level mask for the identification of featured image areas.

\textbf{Training Input}
The source data consist of a single CSV file (comma-separated values), and is thus easily manageable as a spreadsheet with MS Excel or OpenOffice. Table \ref{tab:input-sample} shows an input example.

\begin{table}[t]
  \centering
  \footnotesize
  \caption{Input example}
    \begin{tabular}{lllrrlrrlll}
    \toprule
    method & dataset & split & \multicolumn{1}{l}{epochs} & \multicolumn{1}{l}{segment} & imgaug & \multicolumn{1}{l}{\makecell{batch\\size}} & \multicolumn{1}{l}{\makecell{img\\size}} & \makecell{resize\\filter} & \makecell{color\\space} & \makecell{class\\weights} \\
    \midrule
    VGG16 & \makecell[l]{ISIC-2019} & pre   & 10    & -1 & hflip\_rot24 & 12    & 450   & nearest & RGB   & [0.2,0.8] \\
    SC19  & \makecell[l]{ISIC-2016} & n=100 & 15    & 0.1 &  hflip\_rot4 & 64    & 227   & bilinear & HSV   & compute \\
    \bottomrule
    \end{tabular}%
  \label{tab:input-sample}%
\end{table}%

The \texttt{method} column specifies the CNNs configuration. The \texttt{dataset} column contains the name of another CSV file having a column with the image file names and a second column with ground truth labels. In future versions, we plan to simplify this procedure even more, bringing it to the level of filesystem management, where the input will be a folder whose subfolders represent the classes to predict.
The \texttt{split} column specifies whether the validation and test sets are pre-split on different files (with extra suffix) or alternatively to sample $n$ elements from the training set.
The \texttt{segment} is a float value that, if positive, enables segmentation and specifies the extension factor of the masking area.
The \texttt{epochs} column specifies the number of training epochs, while the \texttt{imgaug} column contains a preset for image augmentation, both affecting training time.
The \texttt{batchsize} column is the training batch size, while \texttt{imgsize} is the (square) resolution at which each input image will be rescaled, both affecting the quantity of GPU RAM needed.
The \texttt{resize filter} specifies the resize sampling strategy (nearest, bilinear, bicubic, or lanczos).
The \texttt{colorspace} specifies whether the images should be kept in their original \texttt{RGB} format or should be converted into {HVS}, {LAB}, or {YCbCr}.
Finally, \texttt{classweights} specifies the weight factors for each class, used as compensation factors in unbalanced datasets. Such weights can also be  computed automatically from the input dataset.

\textbf{Segmentation}
This is the process of automatically detecting the contour separating the lesion from the surrounding skin.
Masking out surrounding skin regions, together with a procedural mask extension, has the potential to improve classification results \cite{burdick_rethinking_2018}. This processing step is optional because there is not guarantee that the segmentation will be correct.\footnote{According to interactive machine learning (IML) goals, we plan to implement the detection of ``anomalies'' (e.g., too many contours in the same image which blocks shape detection, or too small/big areas) to proactively warn and alarm the user to manually correct the pipeline.}

Currently, the segmentation of the input images is performed by a CNN based of the UNET architecture \cite{navab_u-net:_2015} and trained on the ISIC 2017 challenge dataset \cite{codella_skin_2018,tschandl_ham10000_2018}.
On a pixel-by-pixel test, we achieved 73\% sensitivity, 98\% specificity, and a Jaccard Index (aka Intersection over Union, IoU) of 0.69. This compares well with the top results of the ISIC 2017 challenge \cite{yuan_automatic_2019} (83\% sensitivity, 98\% specificity, and 0.76 Jaccard Index).

\textbf{Data Augmentation}
This is the process of procedurally deriving several alterations of an image that look plausible to augment the original dataset. Our image augmentation module has been implemented using a \emph{Decorator} design pattern \cite{gamma_design_1994}.
An abstract \texttt{ImageProvider} class exposes the methods to query for the number of available images and get an image by integer index. A direct concrete subclass \texttt{DiskImageProvider} reads images from files, optionally changing the color space and resizing each image. The \texttt{ImageAugmenter} abstract class (subclass of \texttt{ImageProvider}) provides the base for augmentation. 
Several  augmenters (\texttt{HFlip}, \texttt{Rotation}, \texttt{Brightness}, \texttt{Saturation}) can be concatenated in any order to provide a custom and controlled augmentation chain.

A \emph{Factory Method} \cite{gamma_design_1994} provides a mapping between a mnemonic name and an augmentation configuration.
For example, we are experimenting with three augmentation presets: with \texttt{hflip} every image is flipped horizontally, thus doubling the number of images; with \texttt{hflip\_rot4} every image is flipped and also rotated by 0, 90, 180, and 270 degrees (augmentation 8x); finally, \texttt{hflip\_rot24} (following the schema of \cite{fujisawa_deep-learning-based_2018}) leads to an augmentation factor of 48x.

The image augmentation is performed entirely via CPU, possibly on multiple threads, hence leaving the GPU for the training task only. Whether this is an advantage or not depends on other training parameters.
For example, when training images at 227x227 resolution on a machine with only 4 cores, the augmentation process is actually a bottleneck. However, when training with 450x450 resolution on a 8-core machine, the CPUs are just about 20\% loaded while the GPU is 100\% loaded on training.

\textbf{CNNs}
New CNN architectures can be inserted in the main software by implementing the abstract class \texttt{Classifier} and giving a concrete implementation for the method \texttt{def build\_model() -> keras.Model}.
By design choice, the hyper-parameters of the specific architecture, such as the learning rate, or the type of optimiser and its parameters, are left to the software engineers and hence to the Python code. In future versions, meta leaning frameworks can be added, or AutoML systems that continuously improve over time. 
Only specific pre-sets are visible for the end user.
Again, a \emph{Factory Method} manages the mapping between a mnemonic string and an CNN architecture and some of its parameters.

We are currently experimenting with transfer learning using the \texttt{VGG16} \cite{simonyan_very_2014} network, pre-trained on ImageNet \cite{deng_imagenet:_2009}, on which we substituted the last layers with 2x 2048 fully connected layers and a final softmax. The default optimizer is SGD, but other configurations are available: \texttt{VGG16\_Nadam}, \texttt{VGG16\_Adadelta}, and \texttt{VGG16\_RMSProp}.
Soon, we will perform more tests using \texttt{InceptionV3} \cite{szegedy_rethinking_2016}.
Also, we prepared two non pre-trained networks, one based on VGG16 (\texttt{VGG16\_random}) and the second (\texttt{SC19}) as custom modified version of AlexNet \cite{krizhevsky_imagenet_2012} on the Github distribution. For the feature extraction, we are preparing a CNN based on UNET \cite{navab_u-net:_2015}, trained on the ISIC 2018 challenge \cite{codella_skin_2019} dataset, that is able to extract masks for five features (pigment network, negative network, streaks, milia-like cysts, globules).

\textbf{Training Output}
The output of a training session is written into a directory where a \texttt{train\_output.csv} file contains the same columns of the input file plus a number of columns with the output information, such as the size of validation and test set, class proportions, training time, accuracy, specificity and sensitivity for both validation and test sets, and the ROC AUC for the test set. An additional column is filled with an error message if an exception occurred during the training for the input line (e.g., out of GPU memory).
Additionally, for each input line, the system generates plots for validation and training losses as function of epoch training, together with ROC graphs for both the validation and the test sets.

\textbf{Implementation}
The whole architecture is implemented with the Python language and uses the Keras\footnote{\url{https://keras.io/} -- 23 May 2019} framework (Tensorflow\footnote{\url{https://www.tensorflow.org/}  -- 23 May 2019} backend). All image processing is based on the Pillow\footnote{\url{https://pillow.readthedocs.io/}  -- 23 May 2019} package.
The reference hardware for our experiments is an 8-core i9-9900K CPU, 64GB RAM, and an 11GB nVidia RTX 2080 Ti GPU.

\section{Preliminary results and Lessons Learned}

We ran experiments on the ISIC dataset, as retrieved in February 2019, from which we removed the SONIC subset (whose images contains coloured markers) and the ``2018 JID Editorial Images'' subset (very high resolution, lossless). The resulting datasets counts 12319 images.\\

\textbf{Experiments at 277x277 pixel resolution}
With the VGG16 model, SGD optimiser, images resolution at 277x277 pixels (no cropping, only scaling of the full original image), and image augmentation hflip\_rot24 (48x augmentation), we achieved 0.649 specificity, 0.813 sensitivity, and 0.819 ROC AUC. The training of two epochs lasted about 4 hours and a half. This result is already satisfactory compared to other state-of-the-art approaches like Esteva et al. \cite{esteva_dermatologist-level_2017}, who reached AUC ROC 0.96, but training on a dataset of 120k+ images and augmentation 720x, and like Fujisawa et al. \cite{fujisawa_deep-learning-based_2018}, who reached 0.895, specificity and 0.963 sensitivity but using 1000x1000 pixel resolution images and 24x augmentation.
With all the other optimisers (Nadam, Adadelta, RMSProp) the network was not training properly. More tests are needed to tune the parameters of the optimisers in combination with the other parameters of the pipeline.

\textbf{Experiments without transfer learning}
Since training on a dataset like ImageNet requires months of training on high-end hardware, transfer learning might not be an option when investigating on new architecture, possibly simpler, specialised on the skin lesion domain.
To have a baseline without transfer learning, we trained the \texttt{VGG16\_random} configuration. The network, however, wasn't able to converge after 10 epochs, suggesting that pre-training is not only an option to speed up training, but a necessary condition (at least with this dataset size).

The randomly initialised SC19 architecture showed worse results than VGG16, lacking in sensitivity, with
0.845 specificity, 0.492 sensitivity, and 0.803 ROC AUC. The training lasted about 21  hours for 7 epochs.
However, the loss plot showed a possible overfitting occurring already during the first epoch.
By switching to a lower augmentation policy \texttt{hflip\_rot4} (augmentation 8x), the results improved to 0.814 specificity, 0.674 sensitivity, and 0.835 ROC AUC in 11 epochs, using 1/6th of the computational resources.

\textbf{Experiments at 450x450 pixel resolution}
We increased the images size to 450x450 pixels, which is the lowest resolution available as height of the ISIC images.
Results improved, achieving 0.763 specificity, 0.798 sensitivity, and 0.862 ROC AUC. The drawback is an increased train time of 27 hours for 6 epochs. The same configuration (VGG16, 450px, 48x augmentation) was tested against human performance using the 100-image MClass-D test set \cite{brinker_comparing_2019}, on which dermatologists reached 0.600 specificity, 0.741 sensitivity, and 0.671 ROC AUC.
Our system performed better, reaching 0.762 specificity, 0.850 sensitivity, and 0.846 ROC AUC.
Changing the operating value (malignant vs. benign threshold) from 0.5 to 0.6 leads to 0.862 specificity, 0.750 sensitivity.
This latest results closely match with the performances measured by Brinker et al. themselves with their own CNN on the same test set \cite{brinker_deep_2019}.





\textbf{No impact of image resize filters}
Reducing the size of the ISIC dermoscopic images into 227x227 or 450x450 pixels implies information loss.
We investigated on the impact of the resizing filter over the classification results.
For the following three conditions, VGG16 at 227x227, SC19 at 227x227, and VGG16 at 450x450 pixels, we tried 4 rescaling filters: nearest, bilinear, bicubic, and lanczos (as exposed by the PIL Python package).
Our results show no significant difference in the all the metrics, suggesting that the resizing filter can be left to the default (nearest) for the sake of performances.

\section{Conclusions}
We described  a software toolbox the configuration of deep neural networks in the domain of skin cancer classification. The results suggest that interactive machine learning (IML) design principles should be applied to train effective models in an explorative way. We provided means for the research community to quickly refine skin cancer classification  pipelines by tuning (hyper-)parameters and get feedback as quickly as possible. Interface components need to be simple to end user groups to remain focussed on the machine learning problem at hand. 
The software platform can be used for other (medical) image processing tasks as well, where iterative processes are needed, and users' control on the behaviour of the learning system and latency is sensitive for training. In the future, we investigate the visualisation of image features (aforementioned pigment network, negative network, streaks, milia-like cysts, globules). Alternatively, future work can explore meta leaning frameworks, or AutoML systems that continuously improve over time.




\bibliographystyle{splncs04}
\bibliography{SkinCare}

\begin{thebibliography}{10}
\providecommand{\url}[1]{\texttt{#1}}
\providecommand{\urlprefix}{URL }
\providecommand{\doi}[1]{https://doi.org/#1}

\bibitem{barata_two_2014}
Barata, C., Ruela, M., Francisco, M., Mendonca, T., Marques, J.S.: Two
  {Systems} for the {Detection} of {Melanomas} in {Dermoscopy} {Images} {Using}
  {Texture} and {Color} {Features}. IEEE Systems Journal  \textbf{8}(3),
  965--979 (Sep 2014). \doi{10.1109/JSYST.2013.2271540},
  \url{http://ieeexplore.ieee.org/document/6570764/}

\bibitem{brinker_deep_2019}
Brinker, T.J., Hekler, A., Enk, A.H., et~al.: Deep learning outperformed 136 of
  157 dermatologists in a head-to-head dermoscopic melanoma image
  classification task. European Journal of Cancer  \textbf{113},  47--54 (May
  2019). \doi{10.1016/j.ejca.2019.04.001},
  \url{https://linkinghub.elsevier.com/retrieve/pii/S0959804919302217}

\bibitem{brinker_comparing_2019}
Brinker, T.J., Hekler, A., Hauschild, A., Berking, C., Schilling, B., Enk,
  A.H., Haferkamp, S., Karoglan, A., von Kalle, C., Weichenthal, M., Sattler,
  E., Schadendorf, D., Gaiser, M.R., Klode, J., Utikal, J.S.: Comparing
  artificial intelligence algorithms to 157 {German} dermatologists: the
  melanoma classification benchmark. European Journal of Cancer  \textbf{111},
  30--37 (Apr 2019). \doi{10.1016/j.ejca.2018.12.016},
  \url{https://linkinghub.elsevier.com/retrieve/pii/S0959804918315624}

\bibitem{brinker_skin_2018}
Brinker, T.J., Hekler, A., Utikal, J.S., Grabe, N., Schadendorf, D., Klode, J.,
  Berking, C., Steeb, T., Enk, A.H., von Kalle, C.: Skin {Cancer}
  {Classification} {Using} {Convolutional} {Neural} {Networks}: {Systematic}
  {Review}. Journal of Medical Internet Research  \textbf{20}(10),  e11936 (Oct
  2018). \doi{10.2196/11936}, \url{http://www.jmir.org/2018/10/e11936/}

\bibitem{burdick_rethinking_2018}
Burdick, J., Marques, O., Weinthal, J., Furht, B.: Rethinking {Skin} {Lesion}
  {Segmentation} in a {Convolutional} {Classifier}. Journal of Digital Imaging
  \textbf{31}(4),  435--440 (Aug 2018). \doi{10.1007/s10278-017-0026-y},
  \url{http://link.springer.com/10.1007/s10278-017-0026-y}

\bibitem{celebi_dermoscopy_2019}
Celebi, M.E., Codella, N., Halpern, A.: Dermoscopy {Image} {Analysis}:
  {Overview} and {Future} {Directions}. IEEE Journal of Biomedical and Health
  Informatics  \textbf{23}(2),  474--478 (Mar 2019).
  \doi{10.1109/JBHI.2019.2895803},
  \url{https://ieeexplore.ieee.org/document/8627921/}

\bibitem{codella_skin_2019}
Codella, N., Rotemberg, V., Tschandl, P., Celebi, M.E., Dusza, S., Gutman, D.,
  Helba, B., Kalloo, A., Liopyris, K., Marchetti, M., Kittler, H., Halpern, A.:
  Skin {Lesion} {Analysis} {Toward} {Melanoma} {Detection} 2018: {A}
  {Challenge} {Hosted} by the {International} {Skin} {Imaging} {Collaboration}
  ({ISIC}). arXiv:1902.03368 [cs]  (Feb 2019),
  \url{http://arxiv.org/abs/1902.03368}, arXiv: 1902.03368

\bibitem{codella_skin_2018}
Codella, N.C.F., Gutman, D., Celebi, M.E., Helba, B., Marchetti, M.A., Dusza,
  S.W., Kalloo, A., Liopyris, K., Mishra, N., Kittler, H., Halpern, A.: Skin
  lesion analysis toward melanoma detection: {A} challenge at the 2017
  {International} symposium on biomedical imaging ({ISBI}), hosted by the
  international skin imaging collaboration ({ISIC}). In: 2018 {IEEE} 15th
  {International} {Symposium} on {Biomedical} {Imaging} ({ISBI} 2018). pp.
  168--172. IEEE, Washington, DC (Apr 2018). \doi{10.1109/ISBI.2018.8363547},
  \url{https://ieeexplore.ieee.org/document/8363547/}

\bibitem{deng_imagenet:_2009}
Deng, J., Dong, W., Socher, R., Li, L.J., {Kai Li}, {Li Fei-Fei}: {ImageNet}:
  {A} large-scale hierarchical image database. In: 2009 {IEEE} {Conference} on
  {Computer} {Vision} and {Pattern} {Recognition}. pp. 248--255. IEEE, Miami,
  FL (Jun 2009). \doi{10.1109/CVPR.2009.5206848},
  \url{http://ieeexplore.ieee.org/document/5206848/}

\bibitem{esteva_dermatologist-level_2017}
Esteva, A., Kuprel, B., Novoa, R.A., Ko, J., Swetter, S.M., Blau, H.M., Thrun,
  S.: Dermatologist-level classification of skin cancer with deep neural
  networks. Nature  \textbf{542}, ~115 (Jan 2017),
  \url{https://doi.org/10.1038/nature21056}

\bibitem{fujisawa_deep-learning-based_2018}
Fujisawa, Y., Otomo, Y., Ogata, Y., Nakamura, Y., Fujita, R., Ishitsuka, Y.,
  Watanabe, R., Okiyama, N., Ohara, K., Fujimoto, M.: Deep-learning-based,
  computer-aided classifier developed with a small dataset of clinical images
  surpasses board-certified dermatologists in skin tumour diagnosis. British
  Journal of Dermatology  (Sep 2018). \doi{10.1111/bjd.16924},
  \url{http://doi.wiley.com/10.1111/bjd.16924}

\bibitem{gamma_design_1994}
Gamma, E., Helm, R., Johnson, R., Vlissides, J.: Design patterns: elements of
  reusable object-oriented software. Addison-Wesley (1994)

\bibitem{haenssle_man_2018}
Haenssle, H.A., Fink, C., Schneiderbauer, R., et~al.: Man against machine:
  diagnostic performance of a deep learning convolutional neural network for
  dermoscopic melanoma recognition in comparison to 58 dermatologists. Annals
  of Oncology  \textbf{29}(8),  1836--1842 (Aug 2018).
  \doi{10.1093/annonc/mdy166},
  \url{https://academic.oup.com/annonc/article/29/8/1836/5004443}

\bibitem{krizhevsky_imagenet_2012}
Krizhevsky, A., Sutskever, I., Hinton, G.E.: {ImageNet} {Classification} with
  {Deep} {Convolutional} {Neural} {Networks}. In: Pereira, F., Burges, C.J.C.,
  Bottou, L., Weinberger, K.Q. (eds.) Advances in {Neural} {Information}
  {Processing} {Systems} 25, pp. 1097--1105. Curran Associates, Inc. (2012),
  \url{http://papers.nips.cc/paper/4824-imagenet-classification-with-deep-convolutional-neural-networks.pdf}

\bibitem{masood_computer_2013}
Masood, A., Ali Al-Jumaily, A.: Computer {Aided} {Diagnostic} {Support}
  {System} for {Skin} {Cancer}: {A} {Review} of {Techniques} and {Algorithms}.
  International Journal of Biomedical Imaging  \textbf{2013},  1--22 (2013).
  \doi{10.1155/2013/323268},
  \url{http://www.hindawi.com/journals/ijbi/2013/323268/}

\bibitem{navab_u-net:_2015}
Ronneberger, O., Fischer, P., Brox, T.: U-{Net}: {Convolutional} {Networks} for
  {Biomedical} {Image} {Segmentation}. In: Navab, N., Hornegger, J., Wells,
  W.M., Frangi, A.F. (eds.) Medical {Image} {Computing} and
  {Computer}-{Assisted} {Intervention} – {MICCAI} 2015, vol.~9351, pp.
  234--241. Springer International Publishing, Cham (2015),
  \url{http://link.springer.com/10.1007/978-3-319-24574-4_28}

\bibitem{simonyan_very_2014}
Simonyan, K., Zisserman, A.: Very {Deep} {Convolutional} {Networks} for
  {Large}-{Scale} {Image} {Recognition}. arXiv:1409.1556 [cs]  (Sep 2014),
  \url{http://arxiv.org/abs/1409.1556}, arXiv: 1409.1556

\bibitem{DBLP:journals/corr/abs-1709-01476}
Sonntag, D., Barz, M., Zacharias, J., Stauden, S., Rahmani, V., F{\'{o}}thi,
  {\'{A}}., L{\"{o}}rincz, A.: Fine-tuning deep {CNN} models on specific {MS}
  {COCO} categories. CoRR  \textbf{abs/1709.01476} (2017),
  \url{http://arxiv.org/abs/1709.01476}

\bibitem{szegedy_rethinking_2016}
Szegedy, C., Vanhoucke, V., Ioffe, S., Shlens, J., Wojna, Z.: Rethinking the
  {Inception} {Architecture} for {Computer} {Vision}. In: The {IEEE}
  {Conference} on {Computer} {Vision} and {Pattern} {Recognition} ({CVPR}) (Jun
  2016)

\bibitem{tschandl_ham10000_2018}
Tschandl, P., Rosendahl, C., Kittler, H.: The {HAM}10000 dataset, a large
  collection of multi-source dermatoscopic images of common pigmented skin
  lesions. Scientific Data  \textbf{5},  180161 (Aug 2018),
  \url{https://doi.org/10.1038/sdata.2018.161}

\bibitem{yuan_automatic_2019}
Yuan, Y.: Automatic skin lesion segmentation with fully
  convolutional-deconvolutional networks. IEEE Journal of Biomedical and Health
  Informatics  \textbf{23}(2),  519--526 (Mar 2019).
  \doi{10.1109/JBHI.2017.2787487}, \url{http://arxiv.org/abs/1703.05165},
  arXiv: 1703.05165

\bibitem{DBLP:journals/corr/abs-1803-04818}
Zacharias, J., Barz, M., Sonntag, D.: A survey on deep learning toolkits and
  libraries for intelligent user interfaces. CoRR  \textbf{abs/1803.04818}
  (2018), \url{http://arxiv.org/abs/1803.04818}

\end{thebibliography}

\end{document}